# Stochastic Resonance and Nonequilibrium Dynamic Phase Transition of Ising Spin System Driven by a Joint External Field[*]

Y. Z. Shao[†]   W. R. Zhong   G. M. Lin   J. C. Li

(Institute of Condensed Matter, Department of Physics, Sun Yat-sen University, Guangzhou 510275, China)

**Abstract:** We studied the dynamic response and stochastic resonance of kinetic Ising spin system (ISS), subject to the joint external field of weak sinusoidal modulation and stochastic white-noise, through solving the mean-field equation of motion based on Glauber dynamics. The periodically driven stochastic ISS shows the occurrence of characteristic stochastic resonance as well as nonequilibrium dynamic phase transition (NDPT) when the frequency $\omega$ and amplitude $h_0$ of driving field, the temperature $t$ of the system and noise intensity $D$ attain a specific accordance in quantity. There exist in the system two typical dynamic phases, referred to as dynamic disordered paramagnetic and ordered ferromagnetic phases respectively, corresponding to zero and unit dynamic order parameter. We also figured out the NDPT boundary surface of the system which separates the dynamic paramagnetic and dynamic ferromagnetic phase in the 3D parameter space of $h_0 \sim t \sim D$. An intriguing dynamical ferromagnetic phase with an intermediate order parameter at 0.66 was revealed for the first time in the ISS subject to the perturbation of a joint determinant and stochastic field. Our primary result indicates that the intermediate order dynamical ferromagnetic phase is dynamic metastable in nature and owns a peculiar characteristic in its stability and response to external driving field when compared with fully order dynamic ferromagnetic phase.

## 1. Introduction

The phenomenon stochastic resonance (SR) is of special interest and has aroused considerable concerns of physicists over last decade [1~10]. Contrary to conventional perception of destructive effect caused by noise, in SR process noise probably plays a constructive positive role in transmission of noisy signal and pattern forming through nonlinearly cooperative interaction between random signal and determinant modulation within a system. Generally, the bistable system based upon double well model has been fully studied in comparison with that on Ising model. Some efforts, however, have been devoted to the special studies of SR on Ising model recent years [1,2,4,6~10,12,13]. Among those investigations of SR on Ising model, most of them are associated with the practical application of devices resorting to SR technology. As for a kinetic Ising model, one of topics that are as important and intriguing as SR is about its distinctive feature of nonequilibrum dynamical phase transition (NDPT) induced by an external driving field in this model. Unfortunately, the feature of NPDT on Ising model was investigated mainly under the application of either a determinant driving field, e.g. sinusoidal oscillation, or a stochastic impulsive perturbation [14~20]. To our certain knowledge, there is no systematical study ever reported concerning the feature of NDPT on kinetic Ising model subject to a persistent stochastic perturbation plus a determinant periodic modulation. Focusing on the NDPT of kinetic Ising model driven simultaneously by white noise and sinusoidal oscillation in this paper, we try to give an insight into the dynamic response of Ising spin system to the joint external field as well as the new feature on NDPT owing to stochastic resonance.

---
[*] Project supported by the Natural Science Foundation of Guangdong Province, China (Grant No. 031554)
[†] Correspondence author: stssyz@zsu.edu.cn

## 2. Brief description on nonequilibrium dynamic phase transition and stochastic resonance

2.1 Nonequilibrium dynamic transition of kinetic Ising spin system subject to external field

Nonequilibrium dynamic phase transition occurs within kinetic Ising spin system when a dynamic external field is applied, and the system exists in either dynamic disordered or dynamic ordered state. Considering kinetic Ising spin system with $N$ interactive spins $S$ subject to external field $H(\tau)$, Hamiltonian could be expressed as follow:

$$\hat{H} = -\frac{J}{N}\sum_{(i,j)} \vec{S}_i \cdot \vec{S}_j - H(\tau)\sum_i S_i \tag{1}$$

where symbol $J$ denotes exchange coupling constant and spin $S$ take a value either positive one or negative one. $\sum_{(i,j)}$ and $\sum_i$ stand for the sum of neighboring spin pairs and the sum over all spins, respectively. $H(\tau)$ is a general driving field, including a determinant periodic field (sinusoidal oscillation $h_0 Sin(\omega\tau)$) and a stochastic field (white noise $\xi(\tau)$) as expressed in Eqn.2-1. Symbol $h_0$, $\omega$ and $\tau$ represent the amplitude and the frequency of sinusoidal field, evolution time, respectively. Gaussian white noise conforms to the zero average and auto-correlativity as listed in Eqn.2-2. $D$ is the intensity of white noise and Kronecker Delta function $\delta(\tau)$ is expressed in Eqn.2-3.

$$H(\tau) = h_0 Sin(\omega\tau) + \sqrt{2D}\zeta(\tau) \tag{2-1}$$

$$\langle \xi(\tau) \rangle = 0 \quad \langle \zeta(\tau)\zeta(0) \rangle = 2D\delta(\tau) \tag{2-2}$$

$$\delta(\tau - \tau') = \begin{cases} 1 & \tau=\tau' \\ 0 & \tau\neq\tau' \end{cases} \tag{2-3}$$

For the Ising model with nearest-neighbor interactive spins as described in Eqn.1, we may refer to its mean-field equation of motion based on Glauber dynamics as below [14]

$$\frac{\partial m(\tau)}{\partial \tau} = K\left(Tanh\left[\frac{m(\tau)+H(\tau)}{t}\right] - m(\tau)\right) \tag{3}$$

where static order parameter $m(\tau) = <S_i>$ is ensemble average of all spins. Symbol $t$ and $K$ are the temperature of the system and a phenomenological constant, respectively. For the sake of simplicity, all physical variables in Eqn.3 were reduced to dimensionless form. The dynamic order parameter $Q$ was defined to describe quantitatively the dynamical response of the system to driving field as well as the characteristics of relevant NDPT.

$$Q = \frac{\omega}{2\pi}\oint m(\tau)d\tau \tag{4}$$

The dynamic order parameter $Q$ is defined as the period-averaging of the static order parameter $m$. The two states when $Q = 0$ and $Q \neq 0$ correspond to the dynamic disordered and dynamic ordered state, respectively, as we referred to dynamical paramagnetic phase (P) and dynamical ferromagnetic phase (F) previously. The decrease of the $Q$ parameter from non-zero to zero means the vanishing of dynamic symmetry-breaking and the system evolves into dynamic symmetry (i.e., dynamic disorder) finally while $Q = 0$. Figure 1 exhibits a typical NDPT boundary of kinetic Ising spin system driven simply by a sinusoidal field at zero noise $D = 0$. The NDPT boundary was plotted on the plane of amplitude $h_0$ and temperature $t$, and on the dynamic boundary situated were the critical amplitude $h_{0c}$ and temperature $t_c$ of NDPT. The two regions above and below the boundary in $h_0$~$t$ phase diagram correspond to dynamic paramagnetic phase $Q=0$ and dynamic ferromagnetic phase $Q\neq0$, respectively. The two insets in figure 1 illustrate the time evolution of static order parameter $m(\tau)$ and driving field $H(\tau)$. Static order parameter $m(\tau)$ oscillates symmetrically in accordance with driving field $H(\tau)$ within the region P, resulting in a dynamic symmetry with $Q=0$. In contrast, a dynamic symmetry-breaking with $Q\neq0$ occurs because static order parameter always holds in either positive or negative status, leading to a dynamic ferromagnetic ordering.

The amplitude $h_0$ and the frequency $\omega$ of driving field as well as temperature $t$ are vital to the nonequilibrium dynamic phase transition of kinetic Ising system. When a cooperatively interacting spin system with numerous freedoms of motion is driven by an external field, the thermodynamic response of the system usually lags behind the applied field due to the intrinsic relaxation delay, giving rise to nonequilibrium dynamic hysteresis [17]. This hysteresis is dynamical in origin and disappears in the quasistatic limit. When the time period of the external driving field $2\pi/\omega$ becomes matched in time scale in a certain way with the typical relaxation time $\lambda$ of the thermodynamic system, the dynamic phase arises spontaneously out of dynamically broken symmetries due to the competing time scales in such nonequilibrium driven systems [18, 20]. Studying the match of the two competing time scales in variety of conditions is one of interesting topics on NDPT [15~21].

2.2 Stochastic Resonance

Considering a simple case in which a Brownian particle agitated by fluctuational force, e.g. noise, moves in a symmetric double-well potential. Fluctuational force causes the transition of the particle between the neighboring potential wells with a rate given by the famous Kramers rate [1].

$$\gamma_k=\gamma_0 \mathrm{Exp}(-\Delta V/D) \qquad (5)$$

Prefactor $\gamma_0$ is a positive constant. $\Delta V$ and $D$ are the height of the potential barrier separating the two minima and the noise intensity, respectively. $D$ is in direct proportion to the temperature $T$ as $D=K_B T$. The intensity of noise $D$ has a twofold effect on the jump capability of moving particle over the energy barrier as well as the jump frequency, and this differs significantly from the amplitude of determinant driving field. When stochastic field and determinant periodic driving field applied simultaneously, there are two characteristic time scales in competition: the vibrating period of determinant driving field $\tau_\omega$ and the average transition lifetime of moving particle jumping twice over double-well energy barrier $\tau_k(D)$. Apparently,

$$\tau_\omega=2\pi/\omega \qquad \tau_k(D)=1/\gamma_k \qquad (6)$$

When two competing time scales above match exactly each other in quantity, the original independent stochastic signal and determinant driving modulation interact with each other in synchronization and stochastic resonance arises. It is easy to derive the matching conditions suitable for the occurrence of stochastic resonance as below

$$\tau_\omega=2\,\tau_k(D) \qquad \omega \sim \pi\gamma_k \qquad (7)$$

The match of above conditions could be attained through tuning either the frequency $\omega$ of determinant driving field or the intensity of noise $D$. In this paper, we concentrated on the variation of intensity of noise $D$ and took $D$ as the variable for stochastic resonance. The dynamic order parameter $Q$ defined in Eqn. 4 serves as the characteristic indicator for NDPT resulted from stochastic field. The distinctive intensity of noise $D$ was marked as $D_{SR}$, the characteristic value optimal for stochastic resonance, at which dynamic order parameter $Q$ peaks. Actually, the NDPT owing to stochastic resonance is in nature different from the NDPT caused simply by determinant periodic oscillation.

The statement above makes clear that the three kinds of competing time scales exist in Ising spin system subject to both a determinant driving and stochastic field. They are the typical relaxation time $\lambda$ of the thermodynamic system, the time period of the external driving field $2\pi/\omega$ and the average transition lifetime of moving particle jumping twice over double-well energy barrier $\tau_k(D)$ derived from Kramers rate, respectively. Among the three time scales, the match between $\lambda$ and $2\pi/\omega$ brings about the conventional NDPT while the match of $2\pi/\omega$ and $\tau_k(D)$ leads to stochastic resonance. The new NDPT due to stochastic resonance could be realized if the three competing time scales befit each other in a special way.

## 3. Computational result and analysis

As elaborated above, the dynamic order parameter $Q$ of Ising spin system driven by a external field changes either from zero to nonzero or vice versa when the system undergoes a nonequilibrium dynamic phase transition. We selected the amplitude $h_0$ and the frequency $\omega$ of external field, the intensity of noise $D$ and temperature $t$ as the variables to investigate the dependence of parameter $Q$ on them in this paper. On the consideration of the linear response, the amplitude was limited to $h_0 \leq 1$ while the frequency $\omega$ and the intensity of noise $D$ take values within a wide range. Below are some of typical results.

Figure 2 presents the various curves of reduced $Q/Q_{max}$ versus $D$ at $t=0.8$ and $h_0=0.5$ with different frequencies $\omega$. All curves take on a peak-shape trend and parameter $Q$ attains the maximum at a certain $D$ value denoted as $D_{SR}$, the characteristic feature of stochastic resonance. Only single resonant peak or valley was observed within the wide range of $D$ from $10^{-3}$ to $10^3$. The increase of frequency causes the shift of resonant peak to small $D$, accompanying the change of the shape of curve from peak to valley. The change of the shape of curve above indicates the transformation of the system from a preferentially steady dynamic symmetry to a dominantly steady dynamic symmetry- breaking state.

In addition to the intensity of noise $D$, frequency $\omega$ is another key factor to determine the feature of NDPT and that has been proved by some previous studies without noise [14~20]. The frequency dependence of NDPT under a noise environment, however, remains yet to be studied. In the condition free of noise, the determinant periodic modulation acts as only factor triggering the NDPT of system, and the amplitude and frequency of driving field as well as temperature determine together the dynamic phase region and the exact boundary of NDPT. It is of theoretical significance to investigate the possible change in the feature of NDPT, if any, when a stochastic field is involved. Figure 3 displays the trend of $Q/Q_{max}$ versus both $h_0$ and $D$ at low temperature $t=0.3$, high temperature $t=0.8$ with frequency fixed at $\omega=1$. The stable phase regions of dynamic ferromagnetic ($Q\neq 0$) and paramagnetic phase ($Q=0$) appear within the parameter space of $h_0 \sim D$. At low temperature, the system always stays in a dynamic ferromagnetic state with a terrace appearing at $Q=1$ if both $h_0$ and $D$ keep small ($h_0<0.3$, $D<0.1$), quite similar to the common situation of symmetry-breaking in conventional zero-noise case. When $h_0$ and $D$ take values beyond a larger value ($h_0>0.5$, $D>0.1$), however, $Q$ data surface transforms into a peak shape with $D_{SR}=1$, a new evidence of NDPT featuring on stochastic resonance. In contrast, at high temperature terrace region of common NDPT shrinks substantially while the system-breaking region of stochastic resonance extends as well as the characteristic intensity of noise $D_{SR}$ shifts from original $D_{SR}=1$ at $t=0.3$ to current $D_{SR} \geq 10$ at $t=0.8$.

Figure 4 visualizes the boundary surface of dynamic phase transition in the parameter space of $h_0$, $t$ and $D$ through extruding a noise axis from two-dimensional $h_0$~$t$ boundary as shown in figure 1 by computation. The regions above and below the boundary surface correspond to dynamic symmetric paramagnetic state ($Q=0$) and dynamic symmetry-breaking ferromagnetic state ($Q\neq0$), respectively. The boundary surface neither fluctuates randomly nor declines monotonously with increasing temperature and intensity of noise. The dynamic ferromagnetic phase enhances regularly within some specific regions of $t$ and $D$, as profiled by the boundary surface, implying the dynamic ferromagnetic order induced by noise. The intersection of the boundary surface with the plane $D=0$ is the common boundary of dynamic phase transition as exhibited in figure 1.

In addition to typical dynamic paramagnetic phase ($Q=0$) and dynamic ferromagnetic phase ($Q=1$) existed in kinetic Ising spin system, there appears due to involvement of stochastic field another new intermediate-order dynamic phase (IODP) at $Q/Q_{max}=0.66$ nearly invariable. The region where IODP exists stably is sensitively depended upon other parameters, especially $D$ and $\omega$. We plotted in figure 5 the curve of reduced dynamic order $Q/Q_{max}$ against $h_0$ in variety of noise intensities $D$. The influence of white noise on the transition from dynamic order state $Q=1$ to dynamic disorder state $Q=0$ was demonstrated in figure 5. At zero noise, smooth curve of $Q/Q_{max}$ ~ $h_0$ presents simply a two-state trend, and critical amplitude $h_{0c}$ declines with increasing temperature, as indicated in figure 1. The involvement of white noise brings in not only the fluctuation of original dynamic disorder phase within the range of $Q/Q_{max}=0$ ~ 0.3, but also the presence of the new IODP. Note that the IODP initiates at $Q/Q_{max}=0.66$ first time when $D=0.1$ and the range in which IODP exists stably (the flat extension of $\Delta h_0$ at $Q/Q_{max}=0.66$) extends with increasing $D$ as shown in figure 6a. The temperature dependence of IODP existence, however, is complicated in that two distinctive trends of $\Delta h_0$ versus $t$ were observed at weak and strong noise. Figure 6b indicates that the monotonous drop of $\Delta h_0$ versus $t$ in larger $D$ value turns into a peak-shape trend in small $D$ and the greater the $D$ value is, the faster the $\Delta h_0$ drops. The optimal temperature at which $\Delta h_0$ peaks, if existed, shifts to a low temperature with increasing $D$. The influence of frequency $\omega$ on $\Delta h_0$ is quite prominent as displayed in figure 7. According to figure 7, the flat extension of $\Delta h_0$ comes into view only when frequency $\omega$ takes even number, such as 2 or its double frequency 4 shown in the figure (solid curve), and the extension shrinks once frequency rises. A slight deviation of frequency from 2, e.g. $\omega=1.8$ and $\omega=2.2$ in the figure, causes in consequence the disappearance of the flat extension of $\Delta h_0$ completely. And this very sensitivity to the frequency also discloses the metastable nature of IODP at $Q/Q_{max}=0.66$ in contrast to the stable dynamic ferromagnetic phase at $Q/Q_{max}=1$.

## 4. Summary

We studied the dynamic response and nonequilibrium dynamic phase transition of a kinetic Ising spin system in the circumstance of stochastic resonance. Noise plays a certain constructive role in inducing the dynamic order of kinetic Ising spin system subject to the joint driving of both a stochastic field and a determinant periodic modulation at a proper temperature. Dynamic order parameter peaks at a specific intensity of noise, a typical feature of stochastic resonance. The combinational driving of stochastic field and determinant modulation gives rise to a steady nonequilibrium dynamic phase transition in kinetic Ising spin system when some conditions are satisfied. Similar to the situation of simply a determinant field applied, there exist dynamic disordered phase and dynamic ordered phase due to dynamic symmetry-breaking caused by dynamic phase transition in the case of stochastic resonance. The appropriate matching of three kinds of competing time scales, namely the typical relaxation time $\lambda$ of the thermodynamic system, the time period of the external driving field $2\pi/\omega$ and the average transition lifetime of moving particle jumping twice over double-well energy barrier $\tau_k(D)$, is vital to the occurrence of dynamic phase transition. The variations of the amplitude and the frequency of periodic modulation not only affect the intensity of resonance peak of dynamic order parameter and result in the shift of the peak, but also bring about the change of the shape of curve from peak to valley, indicating the transformation of the system from a preferentially steady dynamic symmetry to a dominantly steady dynamic symmetry- breaking state.

Another peculiar feature on nonequilibrium dynamic phase transition in the circumstance of stochastic resonance is the development of intermediate-order dynamic phase at $Q/Q_{max}=0.66$ in the kinetic spin system. The intermediate-order dynamic phase, in contrast with fully dynamic ordered phase at $Q/Q_{max}=1$, depends strongly upon the amplitude of periodic field, intensity of noise and temperature. Moreover, the intermediate-order dynamic phase also shows a very sensitivity to the frequency of periodic field and only exists at even number frequencies. Judging by the nature of intermediate-order dynamic phase, we regard it as a metastable dynamic ordered phase.

Acknowledgements
This project is financially supported by the Natural Science Foundation of Guangdong Province, China at Grant No. 031554.

Figure Captions

Figure1. The illustrative boundary of dynamic phase transition at zero noise intensity $D=0$. The boundary of dynamic phase transition separating dynamical ordered ferromagnetic ($F$, $Q\neq0$) and disordered paramagnetic ($P$, $Q=0$) phase. Two insets showing the evolution of sinusoidal field ($H$) and magnetization ($M$) versus time within $F$ and $P$ regions, respectively.

Figure2. Dynamic order-parameter $Q$ versus intensity of noise $D$ at various frequencies with a high temperature $t=0.8$ and the amplitude $h_0=0.5$ of driving field.

Figure3. The data surface of dynamic order-parameter $Q$ within the coordinate consisting of noise intensity $D$ and driving-field amplitude $h_0$, while driving-field frequency $\omega=1$ and temperature $t=0.3$, $t=0.8$ respectively.

Figure4. The dynamic boundary surface separating dynamical paramagnetic and dynamical ferromagnetic phase in parameter space consisting of amplitude $h_0$ of driving field, the temperature $t$ of system and noise intensity $D$.

Figure5. The dependence of reduced dynamic order-parameter $Q/Q_{max}$ upon the amplitude $h_0$ of driving field at various temperatures $t$ and different noise intensities $D$.

Figure6. The amplitude extent of driving field $\Delta h_0$, corresponding to a stable intermediate-order dynamic ferromagnetic phase, versus noise intensity $D$ (a) and temperature $t$ (b).

Figure7. The dependence of reduced dynamic order-parameter $Q/Q_{max}$ upon the amplitude $h_0$ of driving field at various frequencies $\omega$ of driving field.

Figure 1

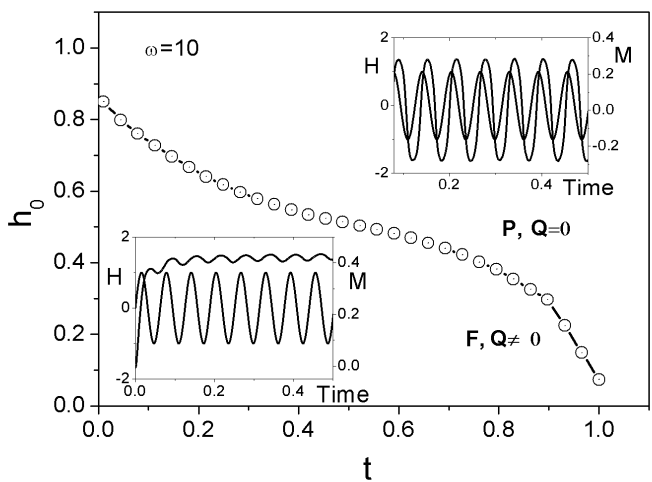

Figure 2

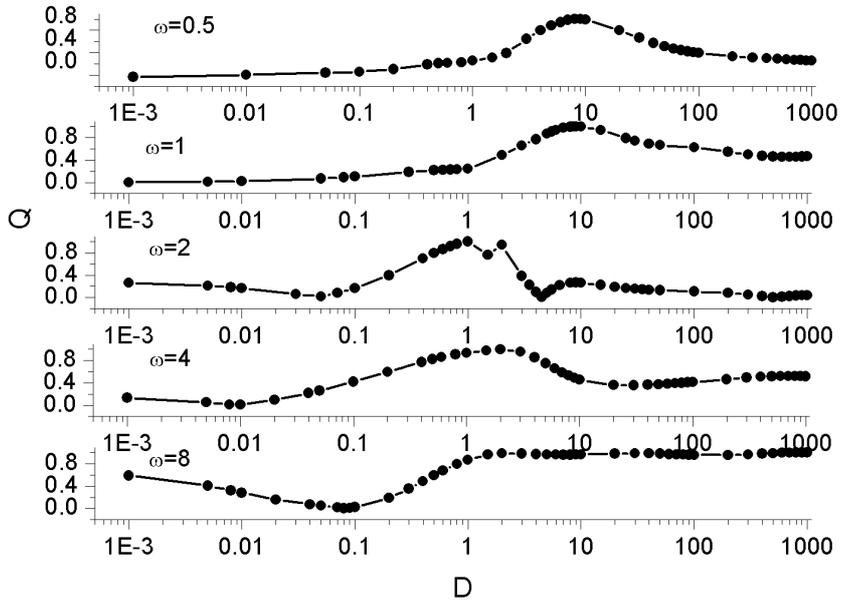

Figure 3

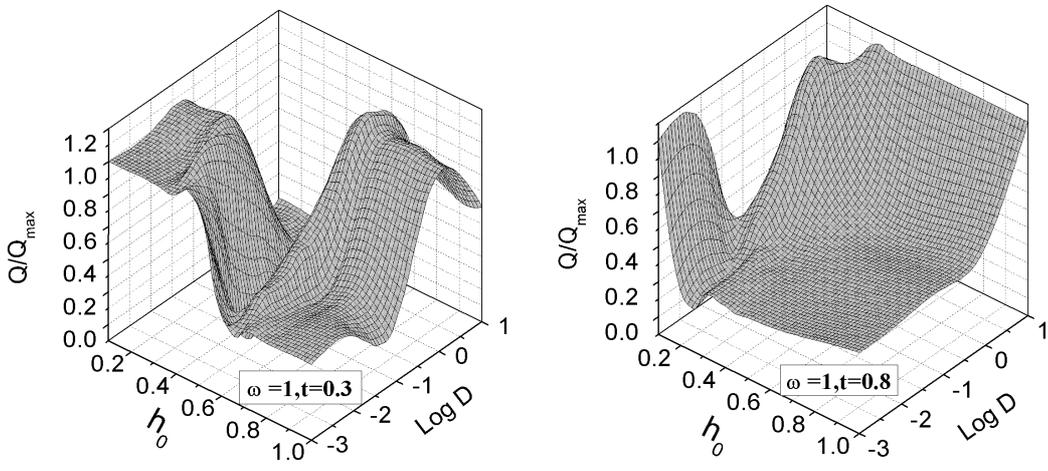

Figure 4

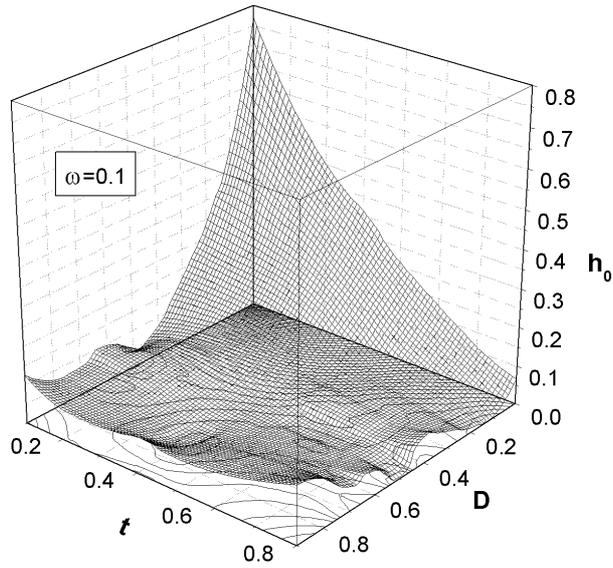

Figure 5

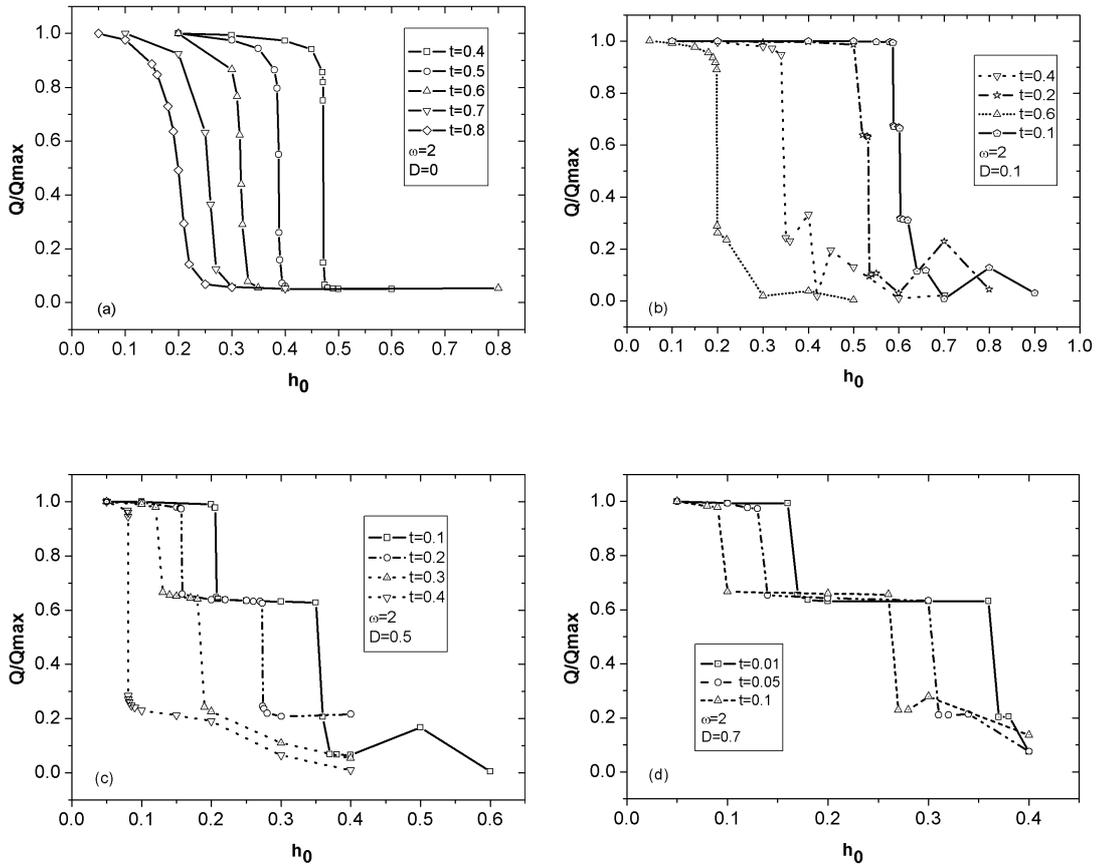

Figure 6

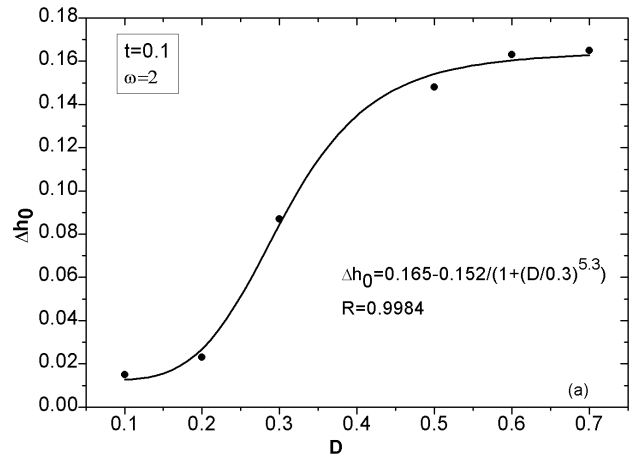

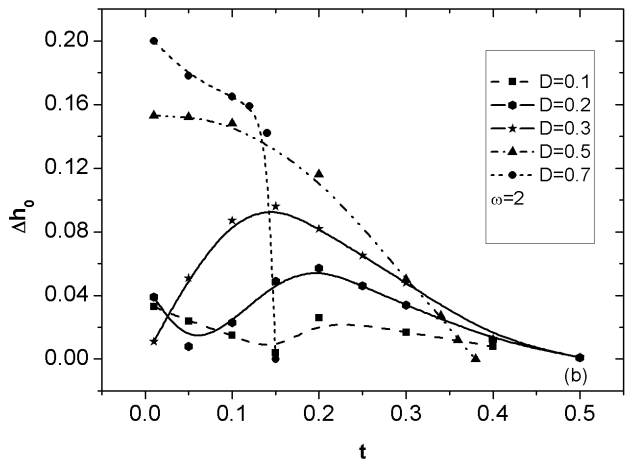

Figure 7

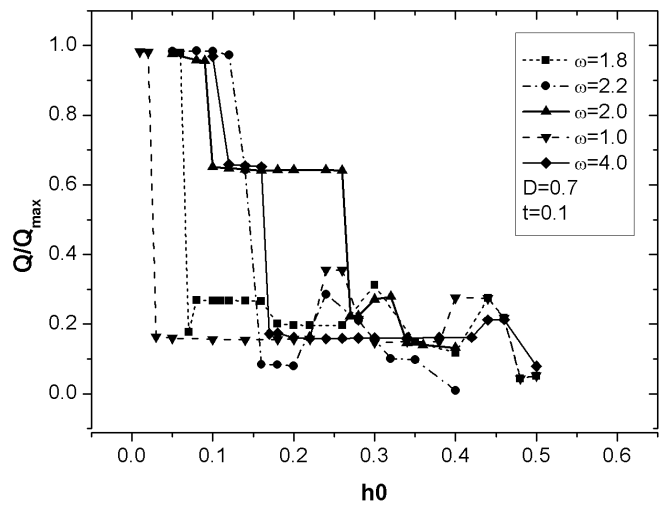